# Dynamical instabilities of two-fluid interfaces in a porous medium: A three-dimensional video imaging study

Prerna Sharma, P. Aswathi, Anit Sane, Shankar Ghosh and S. Bhattacharya

*Department of Condensed Matter Physics and Materials Science,*

*Tata Institute of Fundamental Research,*

*Homi Bhabha Road, Mumbai 400-005, India*

## Abstract

Two-fluid interfaces in porous media, an example of driven disordered systems, were studied by a real time three-dimensional imaging technique with pore scale resolution for a less viscous fluid displacing a more viscous one. With increasing flow rate the interface transforms from flat to fingers and thence to droplets for both drainage and imbibition. The results compare and contrast the effects of randomness, both physical (geometry of the pore space) and chemical (wettability of the fluids), on the dynamical instability and identify the origin of the pore-scale processes that govern them.

A moving two-fluid interface in a porous medium is an archetype of driven disordered elastic media which can exhibit both individual and collective dynamics [1], depending on the competing effects of the elasticity, the randomness due to the disorder and the strength of the drive. Of central conceptual importance is the possibility of the existence of non-equilibrium dynamical phases, of dynamical transitions/crossovers among them and ways of characterizing these phases through the extent of dynamical correlations, both spatial and temporal. Two-dimensional (2-D) porous media serve as model systems which have been extensively studied using experimental [2–7], computational [8, 9] and theoretical methods [10]. However, a realistic porous medium is three-dimensional (3-D) which differs from a 2-D one due to the larger range of accessible pore volumes and greatly enhanced connectivity of the pore space [11, 12]. The more realistic 3-D system is not a simple extension of the 2-D case, either qualitatively or quantitatively as is known for analogous systems [13]. In this letter, we study the phenomenon of displacement of one fluid by another in a fully 3-D porous medium and identify the microscopic mechanisms, dependent on the nature of disorder, that control the macroscopic displacement process. This is achieved by a real time three-dimensional imaging of the interface using a scanning light sheet (SLS) technique [14]. Specifically, we study the downward immiscible displacement of a fluid by a lighter and less viscous one in a model porous medium for variable capillary number (Ca) achieved by changing the flow rate. We compare results for both imbibition, i.e., when a wetting fluid displaces a non-wetting fluid, and drainage which is its converse. The data presented in the paper is mainly for imbibition unless otherwise specified.

A schematic of the experiment with the imaging set-up and the flow arrangement is shown in Fig. 1(a). A model 3-D porous medium was made by randomly packing borosilicate glass spheres (diameter, $D_0$=3mm), in a rectangular glass pipe (hereafter referred to as the "plug") of dimensions $25mm \times 25mm \times 75mm$ mounted vertically. In the experiment, a lipid-rich oil (viscosity, $\eta_1$= 50 mPa.s, density, $\rho_1 = 0.88 gcm^{-3}$ and surface tension with air, $\gamma_1$= 22mN/m) displaces glycerol ($\eta_2$= 780mPas, $\rho_2 = 1.23 gcm^{-3}$, $\gamma_2$ =56mN/m); the viscosity ratio $\mu = \eta_2/\eta_1 \sim 15$. The porous medium is made transparent to light with closely matched refractive indices between the fluids and the solid matrix. A quasi 2-D plane of this transparent medium is illuminated with a light sheet and the fluids containing fluorescent dyes luminesce with characteristic colors while the solid matrix appears dark. The sheet of light is scanned in a direction perpendicular to the propagation of light (see

Supplemental Material for details) and maps the 3-D medium as a set of 2-D image slices (see Supplemental Material, Movie S1).

The left, middle and right panels of Fig. 1(b) show volume-rendered images of the oil in the plug for imbibition at progressively increasing flow rates ($Q$) [15]. With increasing flow rates, the two fluid interface evolves from flat (for $Q = 0.9V_0/s$ where $V_0$ is the volume of a sphere) to fingered (for $Q = 1.7V_0/s$, see middle panel of Fig. 1(b) and Supplemental Material, Movie S2) and finally fragments into droplets (for $Q = 3.1V_0/s$, see right panel of Fig. 1(b) and Supplemental Material, Movies S3, S4). The flat-to-fingered transition at a finite flow rate is the result of a competition between the viscous fingering instability [2, 16] and the stabilizing effect of buoyancy. The critical velocity below which the interface remains stable is directly proportional to Bond number ($Bo$) [17] and inversely related to difference in the viscosities of the two liquids [10, 18]. Even though our experimental system is limited by finite size effects due to the small number of pores in directions perpendicular to the flow, the observed phenomenon is not merely restricted to pore-scale fingers. Indeed, we observe system size spanning macroscopic finger (Fig. 1 (b)) which illustrates collective dynamics. A detailed quantitative analysis of the dynamics of the fingers (see Supplemental Material Fig. S3), shows that aspect ratio of the finger grows monotonically with time, as is expected for an instability.

At higher flow rates, surface tension related effects [19] cause "snap-off"[20] at many pores in the medium resulting in a fragmentation of the interface into droplets[21]. Consequently, the interface itself becomes ill-defined and hence it is not visually clear if the break-up occurs at the interface or behind it. The non-uniform fragmentation yields droplets with a wide range of volumes spanning from many pore volumes to single pore volume, discussed quantitatively ahead. The former represent collective dynamics (correlated motion across macroscopic length scales) while the latter represents individual dynamics (largely local motion independent of one another).

We have observed fragmentation of interface at high flow rates for both drainage and imbibition. The formation of individual droplets (see right panel of Fig. 1(b), Supplemental Material, Movie S3) is likely caused by Roof's snap-off like process at the pore throats [20, 22]. The Roof's static geometric criterion for snap-off to occur in a sinusoidally constricted tube, an example of a non-uniform medium, with wavelength $\lambda$, pore throat radius $r_{min}$ and pore body radius $r_{max}$ is $\lambda > 2\pi\sqrt{r_{min}r_{max}}$[19, 20, 22]. The probable reason for fragmentation in both imbibition and drainage is that viscous effects dominate in both cases at high $Ca$ resulting in substantial thicknesses of the displaced fluid layer due to pore-scale fingering (see Fig. 2(b), (d)), effectively reducing $r_{min}$ and $r_{max}$ and thereby promoting droplet formation. This effect and thus droplet formation is stronger for drainage. Additionally, a 3-D pore network is more prone to droplet formation as compared to a 2-D one [11]. At low flow rates $Q = 1.7V_0/s$, we observe negligible thickness of the displaced fluid layer for both drainage (Fig. 2(a)) and imbibition(Fig. 2(c)) probably due to existence of visually unresolved thin layers of the displaced fluid. Images of the phenomenon of droplet break-up in 3-D are shown in Supplemental Material Fig. S4 (a)-(c). Additionally, we observe coalescence events where two pore-scale fingers form one connected structure (Supplemental Material Fig. S4 (d)-(f)). Together, these pore-scale processes of droplet break-up and their coalescence govern their final size distribution, an important result from the images.

The dynamics of the interface is analyzed using an image subtraction protocol (ISP) which is particularly useful due to the nearly athermal dynamics involved. It consists of a voxel-by-voxel subtraction of the 3-D image of the plug at two successive instants of time such as shown in Fig. 3(a) and (b) for a specific 2-D slice. Figure 3 (c) shows the subtracted image. The white (black) pixels in Fig. 3 (c) represent the glycerol (oil) displaced by oil (glycerol) and the gray ones represent those with unchanged content. As the displacement progresses (at small time ($t$)), the amount of glycerol displaced by oil increases. At long $t$, however, most of the glycerol is either displaced out of the imaging volume or becomes immobile; as a result, their time-incremental displacements decay. Therefore, the time at which glycerol is maximally displaced by oil naturally yields a characteristic time, $T_b$, of the process for a specific flow rate. The time $T_b$ was found to be approximately equal to the time required by the displacing liquid to make a breakthrough across the imaging volume and hence inversely proportional to the flow rate.

Figure 3(d) and (e) show the net volume of those voxels in the entire imaging volume whose content changes from glycerol to oil in a unit time step or vice versa, respectively, as a function of $t/T_b$ for various flow rates. The magnitude of this net volume is governed by the nature of the motion of the two fluids in the imaging volume and is not a direct measure of the efficiency of the process of displacement which will be discussed ahead. Using the ISP, we distinguish three different regimes of flow: (1) at low flow rates, $Q = 1.2V_0/s$ (red circles), the growth of oil in the plug is compact and contents of all voxels change one way, i.e from glycerol to oil;(2) for $Q = 1.7V_0/s$ (green stars) the fluctuations occur both ways, from glycerol to oil and vice versa, and increase upto $T_b$ before decaying to a small value in a comparable time scale;(3) for the highest $Q = 4V_0/s$ (blue triangles) the volume of oil displaced by glycerol does not decay significantly in comparable time scales. The persistence of fluctuations above $T_b$ for case (3) implies that moving oil droplets have reached a quasi-steady state in contrast with the unstable finger.

To further relate the motion of the droplets to the morphology of the porous medium, we prepare a time evolved map of an image slice using what we term a "fluctuation update protocol". Initially, the intensity value of all pixels is set to zero in the map. As the displacement progresses, the value of unity is added to each pixel's intensity each time its fluid content fluctuates, from oil to glycerol or vice versa (Supplemental Material, Movie S5). Thus, the intensity of the pixels is proportional to the number of fluctuations it undergoes. Figure 3(f) shows a color coded map of the resulting intensity of a 2-D slice. The non-uniformity of the map provides an estimate of the spatial heterogeneity of flow. The maxima typically localized in the center of the pore bodies can be readily identified as the primary channels for droplet movement.

Figure 4(a) shows the percentage of the total volume (denoted by the color bar) accounted by droplets of various volumes as a function of the flow rate for imbibition(squares) and drainage(circles). The inset to Fig. 4 (a) shows the contribution of individual dynamics to the morphology by plotting the fraction of net volume accounted by droplets with volume less than the maximum single pore volume $V_p$ in the medium which is $\sim 0.4V_0$ [12]. This fraction was consistently higher for drainage than imbibition and increased with the flow rate. Therefore, the morphology of displacing fluid in imbibition was more compact (see Supplemental Material Fig. S5) due to the capillary forces. For higher flow rates, individual dynamics dominates and the probability distributions of both the size of droplets and the pores exhibit similar characteristic length scales for both processes. These observations suggest that the droplet size distribution [23, 24] is controlled by a snap-off or a Rayleigh-Plateau like instability [19, 20] occurring at the pore scale.

All three types of interfaces, namely, flat, fingered and droplets appear for both drainage and imbibition. The macroscopic property of saturation, given by the fraction of pore space occupied by the displacing fluid, characterizes the efficiency of the displacement process.

The saturation of displacing fluid, $S_{oil}$, computed from the images is shown in Figure 4 (b) for drainage (circles) and imbibition (triangles) for $t \gg T_b$. The saturation progressively reduces as the moving interface becomes more structured. For the highest flow rates accessed in the experiment, the maximum contribution of indvidual dynamics to total volume is at best 20% (see inset to Fig. 4 (a)). Thus, it is the collective dynamics which determines the saturation (see inset to Fig. 4 (b)). Its variation closely resembles that of saturation. It is seen that imbibition aids the displacement only at small velocities where surface tension (capillary forces) are comparable to the viscous effects. At high flow rates, where viscous forces dominate, the wettability of the structure makes little difference in the sweep efficiency.

The study describes the motion of two-fluid interface in driven disordered media in terms of collective and single particle like dynamics with correlation lengths of many pore diameters for the former and comparable, or smaller than single pore volume for the latter. The crossover/transition from collective to single particle dynamics as a function of $Ca$ was found to depend on both physical (geometry of the pore space) and chemical heterogeneity/disorder (wettability of the fluids) in the system. The size of the present experimental system prevents us from quantifying this transition in greater details. Instead, the focus in this paper has been on the identification of micro-scale (pore-scale dynamics) which, nevertheless, affects the morphology of the interface and hence ultimately control the global fluid displacement. Based on the insights herein, one can also investigate the process in more realistic situations of inhomogeneous wettability and more complex pore morphology. This will hopefully yield the long-sought fundamental and predictive understanding of this complex dynamical process that is important for diverse technological applications too.

We thank Amber Krummel and David Weitz (Harvard Univ.), Pabitra Sen and David Johnson (Schlumberger-Doll Research), and D. Dhar(TIFR) for discussions.

---

displacing fluid leaves a thin layer of displaced fluid adhering to the wall due to wettability effects. Therefore, analogy with Roof's process for our experiments is merely to provide a context to gain some understanding.

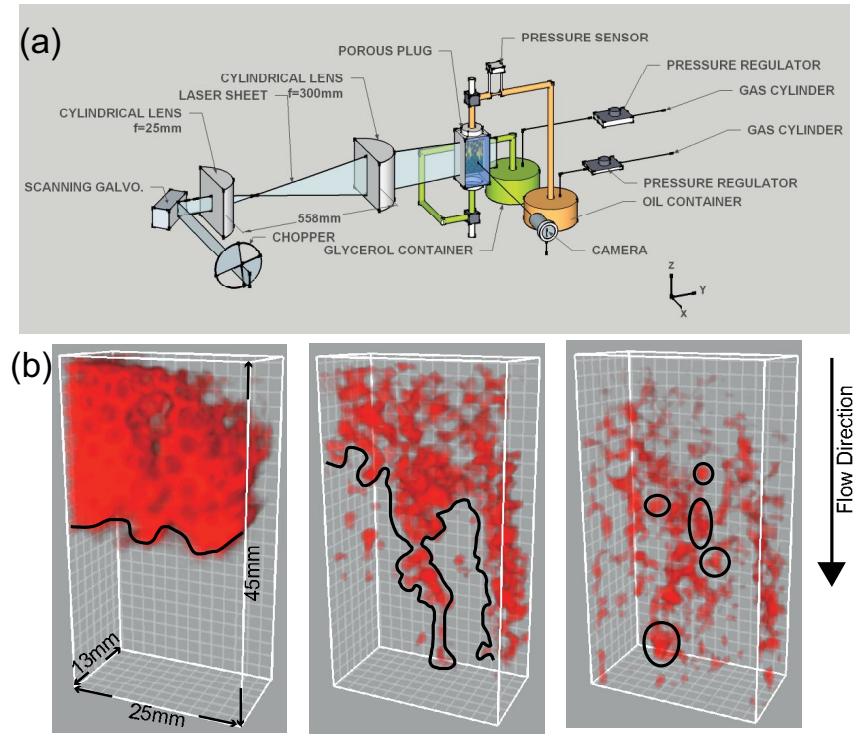

FIG. 1: (a) A schematic of the experimental set-up. (b) The left, middle and the right panel show the volume rendered images of oil in the porous plug corresponding to $Q = 0.9V_0/s$, $Q = 1.7V_0/s$ and $Q = 3.1V_0/s$, respectively.

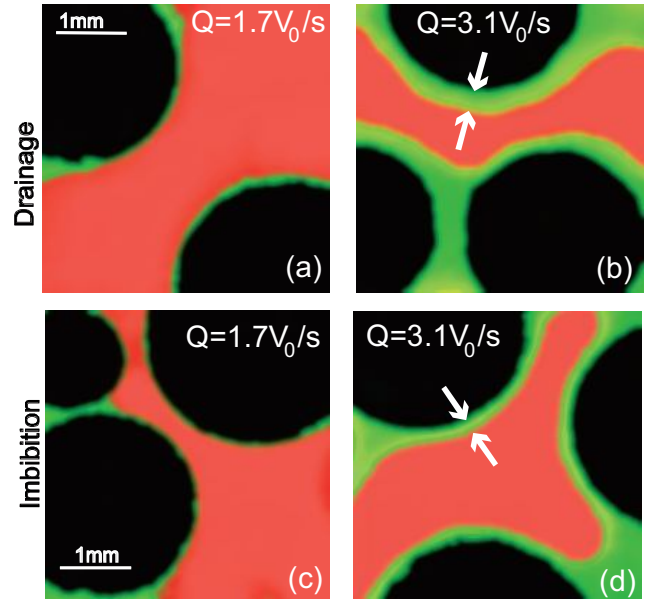

FIG. 2: Images of 2-D slices of the medium shown in top panels ((a) and (b))and bottom panels ((c) and (d)) for drainage and imbibition respectively. The flow rates are specified on the images. Black, pink and green pixels correspond to spheres, displacing oil and displaced glycerol respectively. The thickness of the glycerol layer between the glass spheres and oil (marked by white arrows) is 0.3 mm in (b) and 0.17mm in (d).

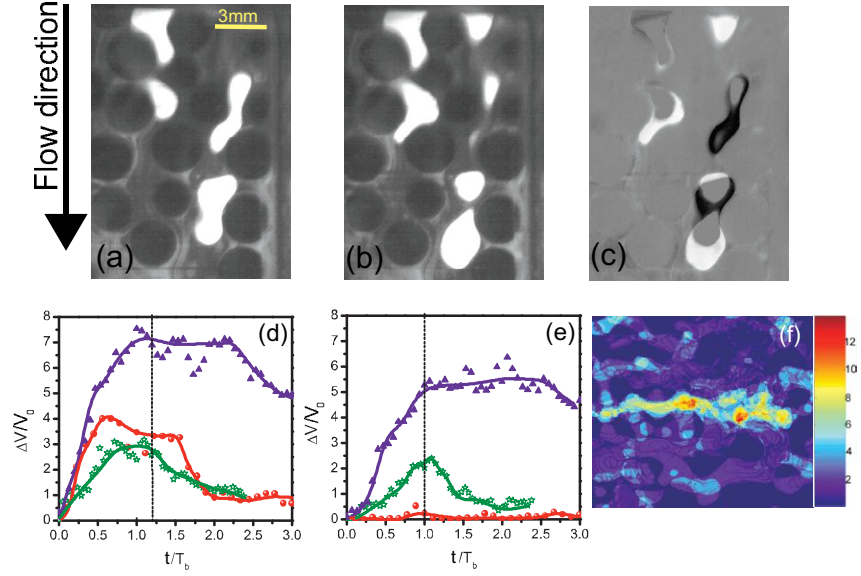

FIG. 3: (a) and (b) show 2-D image slices of a section of the plug at two successive instants of time. White pixels correspond to oil. Black discs are section of spheres while grey pixels correspond to glycerol. (c) The image obtained by the subtraction of image (a) from (b). Glycerol displaced by oil is shown in white pixels and oil displaced by glycerol is shown in black pixels. Gray pixels have their content unchanged between the two time instants. (d) Total volume of glycerol displaced by oil as a function of time for flow rates $Q = 1.2V_0/s$ (red circles), $Q = 1.7V_0/s$ (green stars) and $Q = 4V_0/s$ (blue triangles). (e) Total volume of oil displaced by glycerol as a function of time for flow rates $Q = 1.2V_0/s$ (red circles), $Q = 1.7V_0/s$ (green stars) and $Q = 4V_0/s$ (blue triangles). The curved lines in (d) and (e) are guides to the eye. (f) A time-evolved map (see text for details) for a specific 2-D slice. The color bar on the right denotes the intensity level.

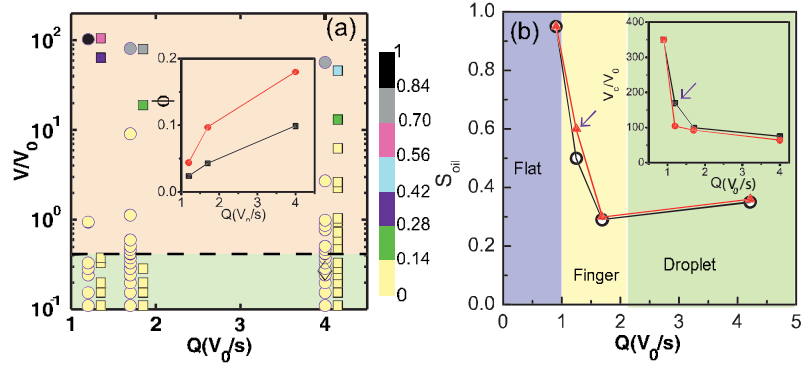

FIG. 4: (a) The percentage of total volume (denoted by the color bar on the right) accounted by droplets of given volume as a function of flow rate for imbibition (squares) and drainage (circles). The data for imbibition has a small offset on the x-axis for clarity. The dashed line denotes the maximum single pore volume of $0.4V_0$ [12]. The inset shows the fraction of total volume accounted by droplets with volume less than $0.4V_0$ for drainage (circles) and imbibition (squares). (b) Saturation, $S_{oil}$, as a function of flow rate for drainage (circles) and imbibition (triangles). The inset shows the amount of volume contributing to collective dynamics (net volume of droplets with volume greater than $0.4V_0$) as a function of flow rate for drainage (circles) and imbibition (squares).

# Supplemental Material

**MATERIALS AND METHODS**

**Scanning laser sheet imaging:** A light sheet of thickness $d_0 = 100\mu m$ ($d_0 \sim D_0/30$) was generated by passing a Gaussian beam from an Ar-ion laser through a set of two cylindrical lenses of focal lengths 25 mm and 300 mm placed 560 mm apart. It illuminated the plug in a plane parallel to the flow direction. As the oil was forced through the plug, the dynamics of the oil front in the porous medium was imaged by scanning the laser beam at 2Hz using a scanning galvo-mirror (GI Lumonics Inc.) and acquiring the images using a high speed monochrome camera (PCO.1200hs) with a speed of 100 frames per second. The camera was triggered by the rising edge of the TTL output of a chopper operating at a frequency of 100 Hz. The images were sequentially stored by the camera on its on-board 4 GB RAM for subsequent processing. The depth of focus of the imaging lens of the camera was large enough to keep the image slices in good focus during scanning. A volume of 25mm X 13mm X 45mm, almost half of the volume of the plug, was imaged with spatial resolutions of $35\mu m/pixel$ and $500\mu m/pixel$ for directions parallel and perpendicular to the flow, respectively.

Fluorescein with excitation and emission wavelengths equal to 488nm and 514nm, respectively, was dissolved in glycerol. Nile blue sulphate, soluble only in non-aqueous oils, had the same excitation wavelength as that of Fluorescein but a different emission wavelength of 560nm in a lipid-rich environment. In a typical experiment, the plug was first flushed with propanol to remove any residual oil. Glycerol was then made to flow from the bottom of the plug against gravity to fill the pore space. Oil, the lighter fluid, was subsequently introduced from the top-end ensuring an initial condition of a flat oil-aqueous interface being introduced in the plug. A differential pressure transducer (Sensym model SX05DN) placed across a glass capillary, connected in series with the plug, of $3mm$ in diameter and $10cm$ in length was used to measure the pressure at which the displacing fluid was being driven.

**Surface treatment of glass spheres to render them hydrophobic:** The drainage experiments were performed with hydrophilic borosilicate glass spheres while imbibition was performed with hydrophobic glass spheres.

The hydrophilic borosilicate glass spheres (Sigma Aldrich) were made hydrophobic by using the silanization protocol as given in [1]. In this protocol, the hydrophilic glass spheres were first baked at $80^0$C for about 12 hours. A solution containing 1 wt% of octadecyltrimethoxysilane (Fluka), 2 wt% of water and 0.2 wt% of hydrochloric acid (37%) in isopropanol was prepared. The baked glass spheres were immersed in the solution and the resultant mixture was continuously stirred for 1 hour. The excess of the solution was removed and the spheres were again heated at $70^0$ C for 2 hours. The contact angle between glycerol and hydrophobic and hydrophilic glass spheres was measured to be $90^0$ and $30^0$ respectively.

**MOVIE CAPTIONS**

The movies are available at http://www.tifr.res.in/∼softmatter/porous.html in QuickTime mov format. The QuickTime player is freely downloadable from http://www.apple.com/quicktime/download/

- Movie S1: The movie shows an image stack made up of various slices obtained from SLS technique. The green and yellow-orange pixels correspond to the glycerol and oil respectively. The spheres were sectioned in a 2-D plane and therefore appear as black (non-fluorescent) discs of various sizes depending on the distance between the slicing plane and a parallel plane spanning a great circle of the sphere. The stack corresponds to a flat oil-glycerol interface formed at $Q = 0.9V_0/s$ for drainage.

- Movie S2: The movie shows time evolution of a finger of oil in three dimensions for $Q = 1.7V_0/s$ (imbibition).

- Movie S3: The movie shows time evolution of dominantly droplet-structured interface in three dimensions for $Q = 4V_0/s$ (imbibition).

- Movie S4: The movie shows an image stack made up of various slices obtained from SLS technique for $Q = 4V_0/s$ (drainage). In contrast with the flat interface shown in Movie S1, oil droplets appear as yellow-orange arbitrary shaped discs in all the 2-D slices.

- Movie S5: The movie illustrates the fluctuation update protocol for a 2-D slice. Each frame of the movie corresponds to the time evolution of the map. The color code shown below reflects the number of fluctuations increasing from black to red.

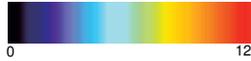

FIG. 1: Color bar for Movie S5.

## CHARACTERIZATION OF PLUG

Figure 2 shows the histogram of the coordination number of the spheres in the porous plug, which is broad and has a peak between 4 and 6, consistent with computer simulations of random packing of monodisperse spheres [2].

## DYNAMICS OF FINGER FORMATION

The fingering instability as shown in Fig. 3(a) is analyzed by estimating the time evolution of the oil content in planes perpendicular to the flow at varying distances from the entry point into the plug (shown as a black cross at the top of Fig 3(a)). The image analysis shows, in Fig. 3(b), that the amount of oil in such a plane increases linearly with time ($\sim t$) and saturates to a finite value in the long time limit. Two time scales are relevant for the problem: (i) when the "tip" of the finger intersects the plane, and (ii) when the plane saturates with oil as the "rear" of the interface intersects the plane. The distances traversed by the tip and the rear, as a function of time, are plotted in the inset of Fig. 3(b), from which the corresponding velocities are obtained. The significantly higher velocity of the tip, $V_{tip} = 0.47 D_0/s$, compared to that of the rear, $V_{rear} = 0.2 D_0/s$, provides a quantitative estimate of the unstable growth of the finger.

The images further show that the growth of the displacing fluid in the direction transverse to the flow is different. It is estimated by analyzing the perimeter of the interface of oil with glycerol and the glass spheres in the planes perpendicular to the flow direction, shown in Fig. 3(c). In the long time limit, the perimeter of the interface, which is also a measure of the surface energy, grows as square root of time ($\sqrt{t}$) in contrast to the linear $t$-dependence of the tip. Different growth rates in the lateral and transverse directions imply that the aspect ratio of the fingers grows monotonically with time.

**ROLE OF WETTABILITY**

In Fig. 5, we compare the terminal structure of oil in the plug for drainage and imbibition at similar flow rates. Figure 5 (a)-(c)(imbibition) and (d)-(f) (drainage) show the volume rendered images of oil in the plug for various flow rates of $Q = 1.2V_0/s$, $Q = 1.7V_0/s$ and $Q = 4V_0/s$, respectively, at $t \gg T_b$.

---

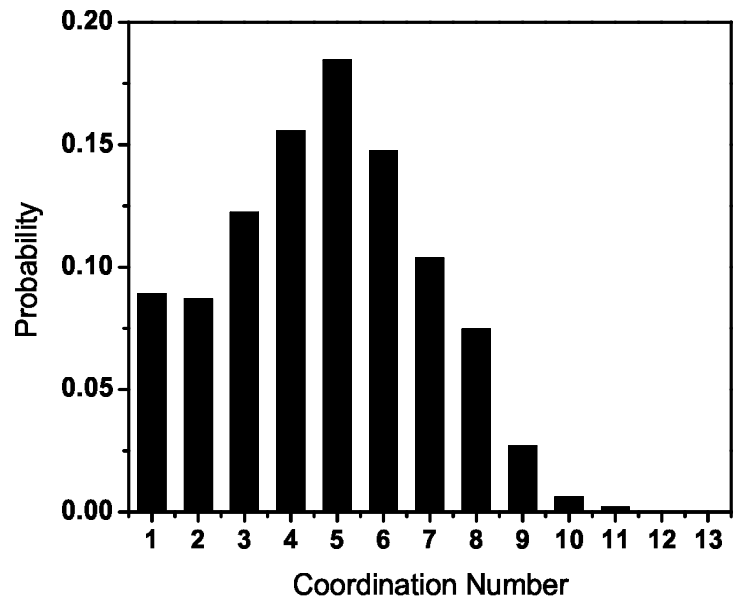

FIG. 2: A histogram of the coordination number of the spheres; the relatively larger values at small coordination numbers is caused by uncompensated finite size effects due to the spheres at the walls.

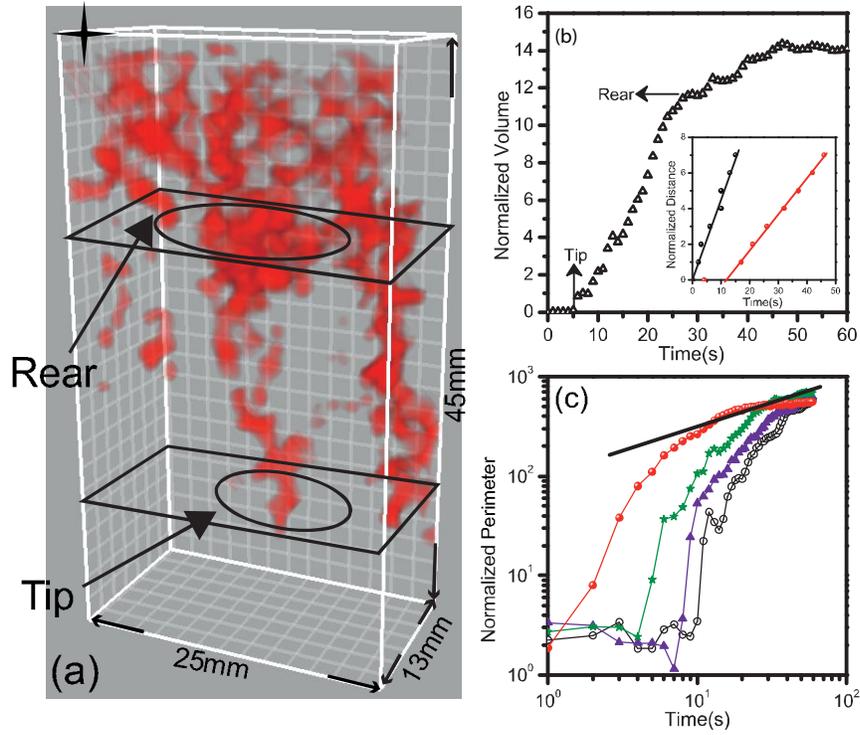

FIG. 3: (a) Volume rendered image of the plug for $Q = 1.7V_0/s$ at $t = 24s$. (b) The computed volume of oil in a plane perpendicular to the flow direction at distance $5D_0$ from the entry point of the plug as a function of time. Inset shows the time instants at which tip (black circles) and rear (red circles) enter a plane as function of distance of the plane (in units of $D_0$) from the entry point of the plug for $Q = 1.7V_0/s$. Solid lines are linear fit to the data. (c) The perimeter of the interface (in units of $D_0$) of oil at different planes perpendicular to the flow direction at distances $3D_0$ (closed circles), $5D_0$ (closed stars), $7D_0$ (closed triangles), $9D_0$ (open circle) as a function of time for $Q = 1.7V_0/s$. Solid black straight line has a slope equal to $1/2$.

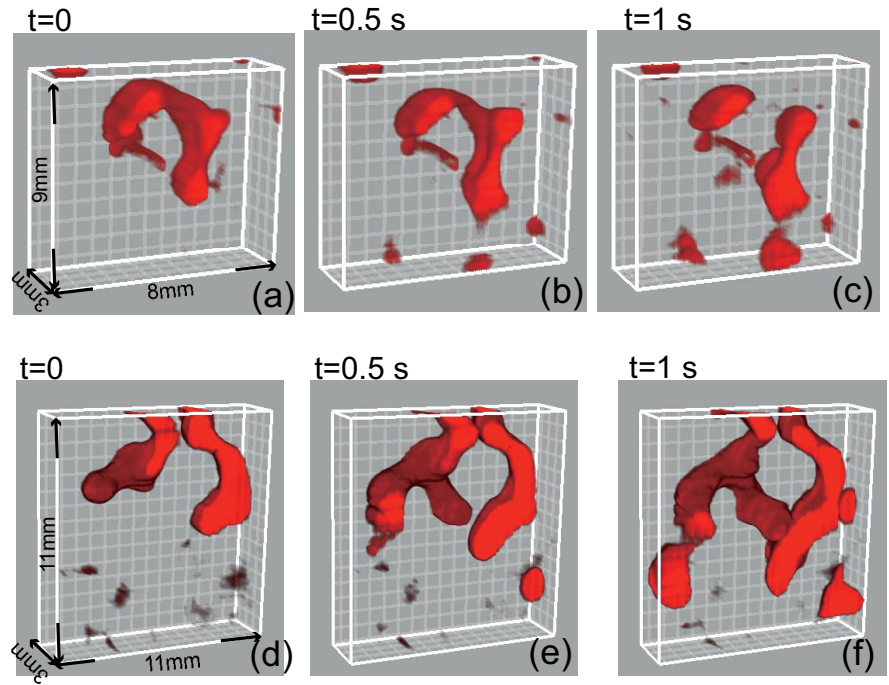

FIG. 4: Panels (a)-(c) show time-lapsed volume rendered images of the break-up of an oil droplet. Panels (d)-(f) show time-lapsed volume rendered images of the coalescence of two fingers of oil.

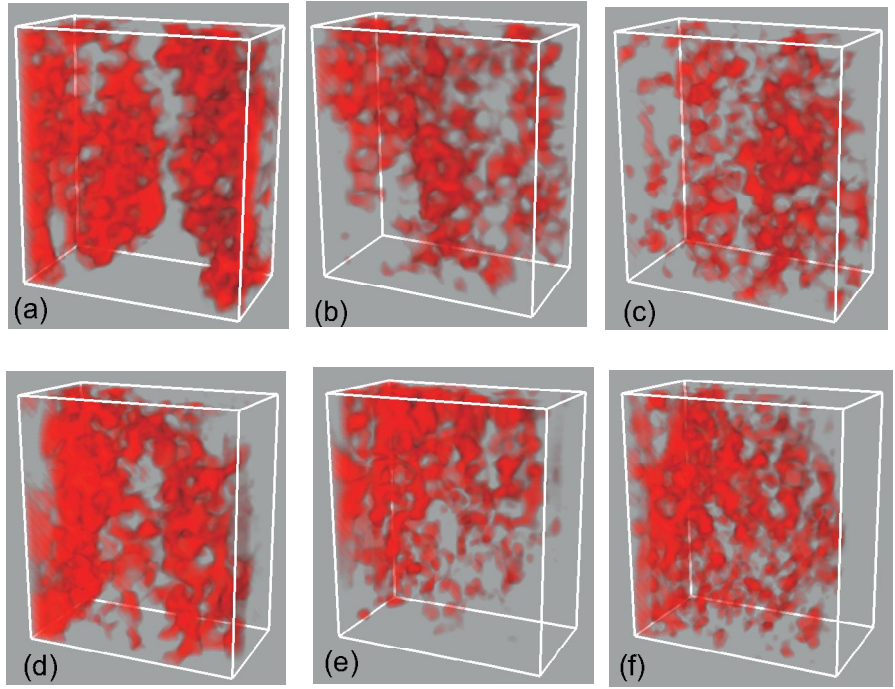

FIG. 5: Volume rendered images of oil in the plug for the flow rates $Q = 1.2V_0/s$, $Q = 1.7V_0/s$ and $Q = 4V_0/s$ for imbibition shown in (a)-(c) and for drainage shown in (d)-(f) respectively at $t \gg T_b$.